# The Management and Use of Social Network Sites in a Government Department

John Rooksby and Ian Sommerville

School of Computer Science

University of St Andrews

Dr John Rooksby

School of Computer Science,

Jack Cole Building,

North Haugh,

St Andrews.

KY16 9SX. UK.

Tel: +44 1334 46 3268

Fax: +44 1334 463278

jrnr@st-andrews.ac.uk

# The Management and Use of Social Network Sites in a Government Department

**Abstract**

In this paper we report findings from a study of social network site use in a UK Government department. We have investigated this from a managerial, organisational perspective. We found at the study site that there are already several social network technologies in use, and that these: misalign with and problematize organisational boundaries; blur boundaries between working and social lives; present differing opportunities for control; have different visibilities; have overlapping functionality with each other and with other information technologies; that they evolve and change over time; and that their uptake is conditioned by existing infrastructure and availability. We find the organisational complexity that social technologies are often hoped to cut across is, in reality, something that shapes their uptake and use. We argue the idea of a single, central social network site for supporting cooperative work within an organisation will hit the same problems as any effort of centralisation in organisations. We argue that while there is still plenty of scope for design and innovation in this area, an important challenge now is in supporting organisations in managing what can best be referred to as a social network site 'ecosystem'.

## 1. Introduction

Social network sites have been widely studied from a consumer perspective. Far less research has been done to address the challenges and opportunities these sites present to organisations. This is despite huge investment by many organisations in social technology, and a number of high profile failures and embarrassments stemming from their use (the media seemingly more keen to report these than success stories). Of the few existing studies of social network site use in organisations, there are a couple looking at the uses of public social network sites (such as Facebook and LinkedIn) in an organisational context (DiMicco and Millen 2007; Skeels and Grudin 2009), and a handful more looking at uses of private, internal sites (Brzozowski 2009; DiMicco et al 2008; Richter and Koch 2008; Romeo 2008).

This paper examines the challenges of managing social network sites in a large organisation. It focuses on the UK Home Office, a department of the UK Civil Service. The Home Office uses social network sites both for internal collaboration and cooperation, and for external engagement with members of the public. At the same time, and in a somewhat contradictory manner, the department has acted to restrict personal uses of public social network sites by its staff. The aim of our paper is not to directly inform or inspire the design of new technology for the Home Office, but to look more broadly and more critically at the ways in which social technologies can be managed and governed by large organisations. A design focus would miss what we would argue to be the key issue facing the Home Office, and presumably other large organisations: there is no lack of social technology; the problem is in how to manage it.

Previous studies of social network site use in organisations have, arguably with the exception of Romeo's (2008) study of Deloitte, focused on large technology organisations (IBM, HP and Microsoft). These organisations seem more tolerant of Facebook and other public social network sites (the use of which is often banned in organisations). These organisations are also developing their own internal social network sites. Other organisations seem to more commonly be procuring commercial systems or services or configuring open source technology (The study of Deloitte (Romeo 2008) for example, examines a Microsoft SharePoint based social network). Our main complaint however is that previous studies, particularly those of internal sites, have focused on a single technology. This misses the wider context in which social network sites will actually be used: people in organisations do not work and communicate across a single technology; they do not work across a single social network site, but across multiple sites and many other ways beside. We are not just saying here that public social network sites (Facebook, LinkedIn, etc.) are used alongside internal ones. We are also saying that large organisations commonly have more than one internal social network site. For example, in a conversation with a member of staff at IBM we were told about more than five internal sites that he was registered with[1]. As another example, in the same workshop that the study of the Deloitte social network site (Romeo 2008) was published, another paper (Kuhn 2008) by a commercial provider mentioned that they manage a (different) social network site for Deloitte[2]. Previous studies have explored aspects of the use of individual social network sites to a depth that our own study cannot. However their focus has guided them away from the broader issues we put forward in this paper. We will look at the management of multiple social network sites. A challenge we will liken to managing an ecosystem.

## 2. Social Technology and Government

Web 2.0 technologies (wikis, blogs, micro-blogs, social network sites, social bookmarking services and so on) are often put forward as a means for addressing, if not overcoming barriers to cooperative work in organisations (e.g. McAfee 2006) including government organisations (e.g. Osimo 2008). Shortly after his election as American president in 2008, Barak Obama caused a ripple of excitement among bloggers interested in web 2.0 when he said:

> "Government should be collaborative. ... Executive departments and agencies should use innovative tools, methods, and systems to cooperate among themselves, across all levels of Government, and with nonprofit organizations, businesses, and individuals in the private sector. Executive departments and agencies should solicit public feedback to assess and improve their level of collaboration and to identify new opportunities for cooperation." (Obama 2008)

---

[1] including BluePages, BeeHive, a microblogging service, a research network, and Connections
[2] The former being for current employees, the latter for keeping contact with alumni

Web 2.0 was not mentioned directly, but many interpreted the phrase "innovative tools, methods, and systems" as pointing towards this technology. Obama, after all, was a president who had drawn upon social media extensively in his election campaign.

We have quoted Obama because he manages to neatly encapsulate the dimensions of collaborative and cooperative work within government: Government departments need to foster cooperative work internally, and across government. They also need to engage with private and non-profit organisations, the public, and other governments (see also Osimo 2008). Clearly these are organisational and communications issues, not strictly technical problems, but it is a reasonable hypothesis that web 2.0 technologies including social network sites can be of benefit. Many, including Osimo 2008, are convinced that these technologies can be used to improve the workings of government.

The use of social technology for interaction with the public is an area in which UK government, by the time of Obama's speech, had already made inroads. Several public consultations had been run over sites such as Facebook, and as we will discuss later in the paper, government campaigns aimed at younger people have been using social media and social network sites for several years. The use of interactive technologies for public facing services and initiatives is an area of growing academic interest. The field of eGovernment or Government 2.0 has emerged to study and support this. The onus of the work here however is on how government organisations can provide web interfaces to their services, how they can bring data into the public domain, how they can run public consultations and so on (Eggers 2005). Studies of social media use are beginning to emerge, particularly in relation to electoral campaigning. The field of Collective Intelligence is also highly relevant. This field has also seen calls for more data to be put into the public domain, and has been promoting the use of social technology as an alternative to the hierarchical forms of decision making in government (e.g. Malone and Klein 2007).

By 2008 the UK government was also developing social technologies for use internally. Rarely, however, in the academic literature are there any detailed considerations of technology and the internal workings of government. Government is a complex area, collaboration and cooperation within public administration is not trivial. The scope and scale of government work requires that it be divided across a large number of departments, agencies, workers and geographic areas and consequently that it be managed in decentralized ways. The interdependencies of this work require, however, that it also be effectively coordinated. Ineffectiveness and inefficiency can occur when decentralised units become self-contained and cut-off from others, a situation popularly referred to in and beyond government as "silo working" (Page 2005). Proposing that social network technologies are to be used to cut across silos is not unreasonable, but this proposal is just the latest in over a century of efforts to improve cooperation and collaboration in government. Efforts to address silo working in UK government over the last century have included: organisational rationalization, moves to market based approaches, the use of coordinating intermediaries, coordination through an active citizenry, and the development of integrated information technology systems (Page 2005). We are discussing new technology but we are not discussing a new problem.

**3. A Study of a UK Government Department**

We have taken a qualitative, investigative approach in this work. Following an extended period of negotiation, our work with the Home Office went ahead between August 2009 and March 2010. The findings in this paper come from a mix of sources: from ten interviews with high-ranking members of the department, from a number of more informal, face-to-face and telephone based discussions with others from the department, from telephone conversations with people from two other government departments, from three site visits during which we were given a desk to work at within the open plan headquarters building, from reading official documents and reports, and the retrieval and analysis of related material posted to various social network sites and other websites. The department is a bureaucratic and security conscious organisation. Our site visits were all escorted, and our interviews observed. Despite this, we were allowed to witness (albeit not closely) a great deal of the mundane workings of the department, we found our interviewees to be candid, and there were no interventions in the questioning (although the observers would sometimes chip in with their own thoughts on a topic). An unexpected consequence of the bureaucracy was that the interim and final reports written during this work were closely scrutinized and commented upon, helping us enormously with the validation of our findings and evaluation of ideas. Report writing became a key aspect of getting this work done, with reports containing a combined total of over 35000 words being submitted and commented upon over the period of research. A final report was finished in April 2010 and commented upon and finalized in June and July 2010.

It is not our intention to give detailed accounts of work with social network sites, and neither is it to present the opinions of our interviewees. Our fieldwork has been used to establish what technologies are in use in the Home Office, what decisions have been made around these, and what the consequences have been. Our interviewees certainly had their own opinions about social network sites, but we have chosen in this paper to give a broad, factual account of what is happening in an organisation. We have been able to cover most aspects of the use of social network sites. This paper does only focus on information that the department classify as public domain (as opposed to protectively marked), some of the details we report about their systems and decision-making have been made less specific at their request, and we have been asked not to dwell on certain, sensitive aspects of their work. Although these constraints under which we have worked have shaped our focus, we do not believe they affect the results or meaningfulness of this work to an academic audience. This work has been funded and undertaken independently of the department.

**4. The Home Office**

The UK Home Office is a large, high profile department within the UK Civil Service. The role of the Civil Service is to deliver public services and develop and implement the policies of the elected government. It is a politically impartial organisation, employed by the Crown rather than Parliament. Around 500,000 people are currently employed within or reporting to the 47 departments of the civil service. The Home Office itself has four main business areas: immigration control, security and counter terrorism, crime reduction, and identity and passport services. It employs 24,000 people and has an annual expenditure in excess of ten billion pounds (16bn US dollars). As might be expected of any organisation of its size (government department or otherwise) the department is made up of several sub-organisations and is dispersed over multiple locations. As also might be expected, the department has been in regular change. It was formed in the late 18th Century and has been reformed many times throughout its history. Currently the department is being heavily affected by budget cuts. These changes will affect the organisation but do not invalidate the themes and results of our study.

The Home Office, as are most departments of the Civil Service, is a large, complicated and changing organisation, and is subject to stringent security demands. The nature of their work makes information sharing and cooperative and collaborative work essential, but the inherent complicatedness of the organisation makes broad scale collaborative and cooperative working difficult. This in turn makes the development and deployment of IT complex. Social technologies may pose a way of cutting across such organisational complexity, but at the same time, these technologies are deployed, used and managed within this complexity.

## 5. Social Network Sites and the Home Office

Social network sites are already being used for collaboration and cooperation within UK government and in their engagements with the public. There are several sites in use: we will mention seven in this paper. Firstly, we will look at two emergent internal social network sites within the Home Office. We will then touch upon the uses and management of public social network sites including Facebook, Twitter, Bebo, and Habbo. These are sites that are or have been used by the department for public communications, but which are restricted for use internally. We will also mention LinkedIn, a public social network site that is not restricted inside the Home Office.

**5.1 Internal Social Network Sites**

Many organisations are deploying their own internal social network sites. At the study site we found not one but two of these emerging. Firstly, the Home Office uses Microsoft SharePoint, an enterprise content management system which has come to have social network functionality built into it and which was increasingly being talked about in the department in terms of its potential for social networking. Microsoft SharePoint is not something we expected from the outset to encompass in our study. It was, however, something that members of the fieldsite were discussing in terms of social networking. Also, upon closer examination of the literature it became clear that at least one other CSCW paper (Romeo 2008) is actually describing a SharePoint based social network site implementation. While SharePoint is not a native of the web 2.0 domain it is apparent that it is a significant technology in the emergence of enterprise social network sites. Networked software products and services such as SharePoint are increasingly being updated and restyled with social networking functionality. Social computing capabilities have been added to SharePoint (and its related administration tools), including profile creation, connections, blogs, wikis, RSS and so on. It is also possible to build upon SharePoint using standard tools and techniques to build social computing environments (this is outlined in (Fu et al 2009)).

A key advantage SharePoint has as a social networking technology at the fieldsite is that it is approved and in existing use. Home office computers are highly managed, it is not possible for users to install new software without permission and difficult to try out new web based services. Therefore existing software packages and services stand the massive advantage that they do not have to be newly installed. This is irrespective of their relative strength, appropriateness or fashionableness for any particular function. The shape of social networking is also rooted within the shape of existing infrastructure. SharePoint has been relied on extensively within the department for a number of years. It was introduced as part of a programme of information infrastructure development and rationalization that also saw the implementation of a corporate file system. The programme went over time and over budget, and was abandoned before the plans to develop standard configurations and practices for SharePoint were implemented. (In what is probably not a totally unusual situation) SharePoint is used in piecemeal and ad hoc ways across the Home Office. Consequently, there is no SharePoint social networking strategy for the whole organisation, but rather these strategies exist among the individual business units that have taken the decision, at a local level, to pursue social networking as a strategy. A business unit that undertakes counter terrorism related work is one such group that is encouraging use of the social network style functionality of SharePoint. One of their key motivations is expertise finding: this group had found they had on occasion been paying for outside expertise on some issues when it actually existed within the group. Therefore they were pushing profile creation in particular. They were also trying to promote blogging and more informal communication within the group. The nature of their work however means that while they increasingly embrace a social networking approach, the potential for wider connection and information sharing will always be limited. Not all groups operated at such high levels of protective marking as the counter terrorism groups, and the upshot of this was that the social network was limited to members of that business unit.

The emergence of social networking style use of SharePoint relies upon, but does not seem to be driven by the availability of social networking functionality in this software. SharePoint seems to be beginning to be thought of differently within the organisation, no longer just a document sharing system, but now that and something to support expertise finding, awareness and informal communication. Our interviewees' reference points for these uses were with systems such as Facebook, and the wider discourses of web 2.0. It was not new technology being taken up here, but rather the emergence of this new technology enabled new ways of seeing existing software.  Seen in the light of social network sites, SharePoint is being thought of and used in new ways.

The second internal site we encountered in our study is a social network for the whole UK Civil Service (i.e. The Home Office and 46 other government departments).  This site, called Civil Pages, is accessible to anyone working within the Government Secure Intranet.  The site is built upon an open source framework, and supports the creation of user profiles, connections to other users, status updates and so on (standard social network site features as recognised by boyd And Ellison (2007)).  Its emphasis, however, is really on group creation, controlled group membership, and document sharing.  The site has not been developed in the Home Office, but in The National Archive, another Civil Service department.  It is an officially sanctioned and centrally funded project.  The Home Office does not have direct control over the site therefore, but can reasonably expect any requirements it has to be taken seriously if not implemented.

Again, this system is shaped by existing infrastructure and the historical context of previous information systems projects and strategies.  Again, there is also a degree of seeing existing technologies in new ways, of seeing them not just for information sharing but for wider social networking style activities.  Civil Pages has grown out of prior initiatives to support interaction between professional groups, and those with common roles and interests across the Civil Service.  The emphasis in these prior initiatives was on the creation of wiki sites.  The software used to manage these wikis was built upon to become the Civil Pages social network, and these wikis were a central aspect of its initial uptake and use.  The process of developing this site started with existing cross departmental forms of interaction and involved both a reframing of how existing infrastructure could be implemented and a reframing of the idea of cross departmental interaction which went from collaborative knowledge production towards informal communication and awareness.

Civil Pages is still fairly new, and during our study appeared to be attracting a great deal of interest but not regular use across the department and wider civil service. Its success as a social network seems far from guaranteed. However, even if it does not become a self sustaining hive of informal communication, it may well continue to transform as an organisational technology. Civil Pages is one of only a few technologies beyond email and paper that allows information sharing between departments across the UK Civil Service. We believe this is significant, and even if it does not become a "Facebook for civil servants" as it is sometimes touted, it is likely to remain an important forum and mechanism for sharing information. There are, however, serious drawbacks and challenges that Civil Pages is posing. Firstly, its functionality clashes with SharePoint. Particularly as a document sharing technology there was beginning to be some clash and confusion. Secondly, individual departments have less control over Civil Pages than SharePoint. They do not have control over what features are implemented, and also do not have control over wider strategy issues. Civil Pages is also funded on a project basis for three years, so there is no guarantee it will be supported over the long term. Thirdly, as an internal social network site, Civil Pages cannot be used to collaborate with people outside of the Civil Service, or civil servants who do not have access to the government intranet. This increases its security, but is also a limitation as government departments do not merely collaborate with each other. For example the Science and Research Group is one that needs to communicate extensively with external individuals, including academics and research organisations. This group still relies extensively on printed documents and couriers. It also employs inspectors who travel between sites and cannot routinely be connected to the secure intranet.

**5.2 Public Social Network Sites**

The other five sites that we will cover are public social network sites: they are run by commercial providers, and are available for use by the general public. As seems to be the norm in large organisations, the use of most public social network sites has been restricted within the Home Office. Access to sites including Facebook, Twitter and Bebo was blocked from computers on the Home Office network. There were several reasons cited by interviewees for this restriction. Firstly, there were concerns in the Home Office about security. Security managers saw social network sites as a serious source of malware. The Home Office had also suffered in the past from breaches to both information security and operational security arising from social network site use. A serious example of this was a widely reported episode in which members of a border security team used Bebo to openly discuss and joke about a case they were investigating[3]. There were also several episodes that were more of an embarrassment to the organisation than a security breach, for example employees using Facebook groups to openly discuss workplace relationships, security breaches, pranks, grievances and so on[4]. The second set of concerns cited in interviews centred on excessive use. Facebook was, before its restriction, the most accessed website from Home Office computers. Because of this, there were concerns about productivity. Their systems provider had also raised the issue that Facebook use was putting unnecessary demand on the network. Thirdly, the Home Office had had a request from the Cabinet Office (a Civil Service department playing a central, coordinative role), asking them to "clarify their position" on social network site use. This message was interpreted as a signal that access should be restricted, and shortly after this message was received the decision to do so was taken. There was no single reason for restricting sites therefore but a combination of factors. There was never any serious investigation or evaluation of what people were doing with social network sites, no questions being asked about what exactly was being done on Facebook etc. Most of our interviewees were not social network site users, and in this situation it seems the only visible uses of Facebook to senior managers in the department were those that led to embarrassments and security breaches. The restriction covered all sites but one, LinkedIn. This was not blocked on the rationale that it is a 'professional' social networking site. Again, there has not been an investigation of what specifically is done by employees with this site.

---

[3] "Immigration Officials Suspended Over Racist Internet Jokes." The (Scottish) Daily Record, Feb 11 2008.
[4] "Two Faced. CIVIL SERVANTS reveal how they SKIVE OFF work and MOCK immigrants — in a flood of astonishing messages on chat website Facebook." The News of The World, June 12 2009.

This restriction is far from the end of the story for public social network sites. We have found restriction serves to change the shape of social network site use, not to stop it. The restriction is enforced by blocking access to the sites from Home Office computers, and as such it does not affect mobile devices, home computers and so on. Many people within the department have mobile phones with internet access (often as their own personal device). There was no reason why people could not 'waste' time accessing Facebook this way during the working day. Most of the department's employees are also likely to use Facebook at home (or elsewhere in their leisure time). Indeed, it struck us that the problems the department had faced from inappropriate materials being posted to social network sites was likely to have involved, if not have originated from personal computers. Certainly a noticeable number of people who have posted to Home Office related Facebook groups do not actually work for the department, but are alumni reminiscing about their time working there. It struck us that restriction did not solve the problems that were seen to arise from social network site use, and at worst simply made it much more difficult to police the ways in which sites were being used. It also caused new problems, as we will discuss.

Despite the restrictions, members of the department continue to make use public social network sites in official and unofficial capacities. Primarily, a business unit has been using public social network sites in anti drug and anti knife-crime campaigns. Both are long running, multi-million pound, cross medium (TV, web, and print) campaigns. The anti drug campaign "Talk to Frank" encourages people of all ages to access information about drugs and talk with advisors. It has previously used the social network site Habbo over a one-month period to provide a way for people to talk to advisors. Habbo was seen as providing a mechanism for engaging with young people. More recently, Facebook has been used, with a page being created for the character "Pablo the Drug Mule Dog" who simultaneously featured in television adverts. This Facebook page ran for several months and gained around 200,000 fan connections, before this aspect of the campaign stopped and the page deleted in May 2010. The "It Doesn't Have to Happen" campaign is targeted specifically at 10-16 year olds and encourages them not to carry knives. This campaign uses the social network site Bebo, with their page attracting over 10,000 connections. Bebo was at the time of the campaign launch extremely popular among the target audience, although has since lost a lot of its market share to Facebook. Recently the very survival of Bebo has been called into question. The department has had to make judgements about which sites to invest their own efforts in, but the populations of these sites and the very sites themselves are not necessarily stable. The department can make informed decisions, but has no actual control here. They also cannot control the fact that it is easy to spoof and imitate campaign sites, for example with fake profiles for Pablo the Drug Mule Dog being created. These campaigns were begun before the restriction of social network site use in the Home Office were implemented. They were severely affected, when the decision was taken to restrict access to social network sites, their use in these campaigns was overlooked. To work around the restriction, 'standalone PCs' not on the department intranet had to be sited and used. This allowed the campaign to continue but complicated opportunities for monitoring them and for copying and pasting information from documents to the sites. It also meant that extra room had to be found for new computers within the existing office space (something far less trivial than might be assumed).

Another, public facing use of social network sites by the Home Office is to distribute announcements via Twitter. Unlike the campaigns described above, which use named people or characters and do not mention the Home Office, the Twitter account is in the name of the Home Office. The account is predominantly used to announce links to news items on their website, press releases, or video and images on YouTube and Flickr. The Home Office Twitter policy, as stated on their website, is not to reply to messages. They state however their intention is that any messages to them will be collected and forwarded within the department to relevant people as necessary. Clearly Twitter is in no way a conversational tool for them, and given the difficulties of copying and pasting between standalone machines from which Twitter is available to ones on the intranet it is unlikely that messages to them are speedily forwarded. An increasing number of UK government departments are using Twitter (Williams 2008), but the bureaucratic, information security conscious ways in which departments operate, together with the fact that access to this site is restricted, mean that it can only practically be used to broadcast announcements rather than to interact with people. These non-interactional, somewhat unsocial uses of Twitter, are accounted for by (almost excused by) the Twitter policy written on the website.

A more inward use for social network sites by the press and communications units concerns their monitoring for news and announcements. Both units have seen their focus shift extensively from printed to digital media over the past decade, and more recently they have had to deal with social network sites. We were told the press office is regularly contacted regarding stories that have originated on social network sites, and both the press office and communications directorate need to have some awareness of what is going on. Both saw access to social network sites, especially Twitter, as providing useful information about current concerns. The restrictions made this difficult, and again standalone PCs had to be used as a workaround.

Other uses of public social network sites include use for employee networking and support. For example an official LGBT (Lesbian, Gay, Bisexual, Transgender) group within the department, whose role includes providing peer support, as well as some more watchdog activities, uses a Facebook Group. Other, more informal groups were also using Facebook, for example to arrange jogging, socializing and so on. The main use of Facebook and other sites must simply be employees using them to communicate directly between themselves and with others. Work and social lives are often blurred, and most social network site users are likely to have work colleagues among their connections. What strikes us from our study is that these kinds of interaction only seem to become visible to organisations when they lead to embarrassment. Managers were not using these sites, and there seemed to be inadequate awareness of the role they played both officially and unofficially. Although it had made some major decisions about them, the Home Office had not made efforts to understand social network site use among its staff. (Such a study is something we have proposed as future work, and is not reported in this paper).

The blanket restriction of sites was being called into question at the time of our study. Certainly the effects of the restriction on the official uses of social network sites for public engagement were recognised to be damaging. An option being seriously considered was whether selected people could be given access if they are able to demonstrate a business purpose for using the site. This sort of selected access posed some complexities in implementation, both in terms of technical changes to the way the network is managed but also administrative changes. In particular a process to evaluate business purposes for using the sites would need to be created, and a way of tracking permissions needed.

Briefly, we should mention the department arguably has in certain, limited respects, world leading expertise in social network site use and technology. This is related firstly to child protection and secondly to surveillance. The department holds responsibilities related to child protection in the UK, including the protection of under 18 year olds on social network sites and have issued advice about children's safety online, lobbied social network sites on adding child protection features, and have developed a so called 'panic button' for Bebo and Facebook. The actual work on this is undertaken by one of the Non Departmental Public Bodies (quasi-independent, "arms length" bodies) reporting to the department, but it is apparent that the department can be supplying (high quality) advice and software for certain social network site users, and yet have a much more confused approach at the level of their own business. There is a similar issue regarding surveillance. Our study did not address what capability or expertise the Home Office has on internet surveillance, but it is public knowledge that the department has in the past been working towards creating a legal framework and methods for monitoring the uses of Facebook and other social network sites (but to our knowledge has never progressed very far with this). Our point is that even if you have some expertise in social network site technology and use, this does not necessarily entail expertise in its use in the context of work, or necessarily diffuse into organisational policy and decision-making.

## 6. The management of social network sites

The major challenge regarding workplace social networking technology appears not to be how to design an ideal system, but how to manage existing and emerging systems. Our account of social network site management and use in the Home Office has touched upon seven technologies. Five major problems of managing these are given below.

### 6.1 Boundary Problems

Organisations such as the Home Office face two kinds of boundary problem. Firstly, they face problems associated with organisational boundaries. For them, there can be no such thing as social networking "inside" and "outside" the organisation. Such a simple distinction is not possible because organisations in themselves have no simple inside or outside and their boundaries are not hard and fast. The Home Office houses several agencies (including several intended to be "arms length bodies"), and is itself part of the Civil Service. There are also many different groups working with different levels of security clearance. There too are people who work closely with the Home Office but are not government employees. Creating an internal site requires deciding what "internal" means.

Secondly, they face problems associated with boundaries between work and social life. The Home Office was concerned that employees were using public social network sites for personal reasons during the working day, that employees were spending time socialising and communicating about issues unrelated to work. It was also worried about the use of external social network sites for the reporting of internal socialising and public discussions of confidential information, with sometimes sensitive and often embarrassing information being placed into the public domain. However, the department was arguably too focused on negative aspects of social technology use. The use of sites for organising social gatherings among employees is arguably beneficial to both the exchange of information and to job satisfaction (Skeels and Grudin 2009). Sites were also being used to manage existing troubles that extend across personal and working boundaries, for example through their use by LGBT and disability groups.

**6.2 Limited Control**

The department sought to control the use of social network sites, an approach that proved counter productive. The restriction of access to external public social network sites was mistakenly seen as being a method of controlling their use. Such restrictions actually greatly reduced the ability to monitor the uses of these sites, which will simply continue over different channels. Restriction also creates complexities for the use of social network sites for public engagement, it requiring new administrative frameworks, new computers to be sited, and it complicating moving information between these sites and internal systems. The use of these sites is near impossible to control, and instead appropriate ways to manage them need to be devised.

Control problems relate not just to external sites but to internal ones as well. At the Home Office, SharePoint use is something that could, in theory, be systematically managed at a departmental level. However, the project that would have put the appropriate facilities in place for doing this was abandoned before completion. Instead SharePoint was being managed more on a group-by-group basis. Civil Pages on the other hand was controlled from a different department within the Civil Service. The management and design of Civil Pages is something that the department can have a say in, but for which government-wide requirements would have to be taken into account. With IT in the Home Office usually managed at a departmental level, Civil Pages is actually quite a challenging technology. It requires new ways of working in IT management.  Social network sites in general, sit very uncomfortably with command and control style management.

**6.3 Visibility Issues**

It became apparent during our studies that among the managers of the department there was limited visibility of how social network sites were being used.  Stories in the press seem to have been the core form of insight into the use of external sites, and these arguably painted an unduly negative picture of the effects of these sites on the workings of the department.  Perhaps more could have been done to monitor use, but the connections between people and the information they share is not easy to see on a broad scale, particularly on external sites.  The decision to restrict access to external sites compounded this problem by reducing the ability to check activities on Facebook group pages and so on.  Internal sites present greater opportunities for monitoring, for example the uses of Civil Pages, who is registered, what groups there are and so on, could be roughly seen by anyone in the Civil Service logging into it.  SharePoint's visibility on the other hand is quite clustered, the fact that one group uses it intensively may not be visible to people outside this group.

The uses of social network sites in an organisation are diverse and difficult to monitor and comprehend.  The readily available pictures of use are unlikely to be comprehensive and sometimes can be misleading.  For an organisation to understand what is going on, we believe qualitative research including surveys and interviews will be necessary.

**6.4 Overlapping functionality**

The functionalities of social network sites overlap with each other and with the functionalities of other technologies.  Internal social network sites support document and information sharing, but this overlaps with existing technologies available for this purpose (e.g. groupware technologies and wiki sites).  Internal and external sites also present new methods for communication, but ones that will overlap with existing means (email and telephone for example). Social network sites are not supporting activities that are wholly new or separate to existing activities, and so the relationship between social network sites and other technologies needs to be managed.

**6.5 On-Going Change**

Social technology is an unstable domain. New, external social network sites are launched regularly, with some gaining short term popularity and others seeing more sustained use. Individual sites are also changing, with features regularly added or removed. According to Lampe et al 2008, not just the technologies, but practices of social network site use are changing over time. This seems to be true at the Home Office. Facebook groups for example, which were at the heart of several embarrassing episodes for the Home Office, seem to be seeing significantly less than they were only a few years ago.

Internal sites are also emerging and evolving, but it should be pointed out the way these emerge is firmly rooted in existing infrastructure. SharePoint is an existing feature of Home Office infrastructure, and as a result a default choice for expanding social network activities. The government operates a secure network in which it is not possible simply to download and play with new software, or even to access a lot of web based software services. Similarly, Civil Pages grew out of an existing system used originally to manage wiki sites. We have mentioned in this paper that the emergence of social network technologies from existing groupware systems involved in part a reconceptualization of existing technology, civil servants began to see and talk about technologies such as SharePoint in terms of social networking.

Constant change means social networking is an area for which it is difficult to develop long-term strategies. The department may chose to invest in internal sites such as Civil Pages, but this technology is funded on a project basis over three years which means that the technology is not only out of their control but that the continuation of this system is not guaranteed. The department also invested effort in some public social network sites such as Bebo that over time saw falling use by the public. Long-term strategies may be identified that involve social networking, but it is risky to tie this into to any particular platform.

**7 Discussion: Social Network Sites and Collective intelligence in Organisations**

We have described the issues and problems faced by a government department with respect to social network sites. Descriptive or "scenic" (Button 2000) approaches are sometimes criticised but (perhaps because we were working with managers) in this case we have found them to be valuable. The department needed to take stock of what was happening with social technologies in order to work out exactly what the problems were and how they might be coherently addressed. More generally, we believe a descriptive approach is necessary because research is not currently addressing all the problems faced by large organisations. We believe research is currently too focused on designing individual technologies, and is overlooking organisational complexity. To explore this point, we will discuss research into "collective intelligence". This is just one area of several exploring the opportunities social and collaborative technologies bring. We feel this area is particularly relevant to our discussion as its vision extends to new forms of collaboration across and beyond government.

Earlier, we reported Obama's assertion "government should be more collaborative". A much more extreme view on this issue has been put forward by Thomas J. Malone, director of the Center for Collective Intelligence at MIT. At the opening of the Center, in 2006, Malone posed the key question for his field as: how can people and computers be connected so that collectively they act more intelligently than any individual, group, or computer has ever done before? Answers to this question, claimed Malone, could pave the way for:

> "… better ways to organize businesses, to conduct science, to run governments, and--perhaps most importantly--to help solve the problems we face as society and as a planet." (Malone 2006).

In a paper on addressing climate change, Malone and Klein (2007) propose that there needs to be a large scale, massively participatory debate. Politicians, civil servants, scientists and the general public should share their information freely and present their reasoning in a logical and accountable way (similarly McGovern (2010) argues that the global financial crisis can best be solved through collectively intelligent approaches, and Schuler (2001) argues a case for "civic intelligence"). Our concern with such visions is not whether democratising information and decision making to an extreme is appropriate, but with whether this can be achieved through software design. Too often it is assumed that the vision of collective intelligence can be achieved through a single technology, for example Malone and Klein (2007) proposes that civil servants, scientists, activists and the public collaborate through an online argumentation system.

Other collective intelligence research looks more specifically at how this can be fostered within enterprises. Smith (1994) and others have suggested methods for designing groupware that would foster collective intelligence. Convertino et al (2010) have suggested that existing technologies such as MS SharePoint and IBM Beehive should be considered as collective intelligence technologies. What we want to point out is that any new social technology deployed will become one of several others in use, and will likely be used for purposes beyond supporting collaborative work. Ultimately the technologies will be subject to the complexities that it is hoped they will cut across. We suggest the problem that organisations face if they want to foster greater collaboration is not how to comprehend, design or procure individual social technologies but how to create appropriate strategies for working and managing across technologies. The field of collective intelligence, if it is to transform government or any large organisation, will need to address organisational complexity. It will need to address not just what features of a social network technology are desirable in an organisation, but how these can be deployed, integrated and managed in a real setting.

We suggest the problems of managing social technologies can be thought of as the problems of managing an ecosystem. The ecosystem we have described features a number of interdependent technologies that are difficult to observe and which stretch outward beyond the organisation. As such the ecosystem presents limited opportunities for control, and the introduction and modification of technologies both inside and outside of the organisation can have unanticipated consequences. The success of any technology is also dependent upon factors in the wider ecosystem. We do not mean the organisation is powerless or unable to act in the face of social network sites. Rather we mean that organisations can only seek to tame but not control the ways in which technologies are being used. The problem for organisations such as the Home Office in improving collaboration is not to supply the right technology but to support and enable organisational members in their uses and choices about technology.

## 8. Conclusion

We have found that the UK Home Office, a department of the UK Civil Service, is facing challenges in managing the use by its employees of multiple social network sites. These sites are overlapping, conflicting, evolving, rooted in existing technology and infrastructure and are embedded within organisational procedures and demands. Ultimately they are rooted within the very complexities they are often pitched as cutting across. The idea that there can be separate workplace and leisure social network technologies, and even the idea that there can be a single, central social network site for an organisation has not stood up to scrutiny. It may be that some organisations are able to separate and compartmentalise the uses of social technologies better than our fieldsite, but we believe the story we have told here will not be unusual for large organisations.